# Global Financial Cycle, Commodity Terms of Trade and Financial Spreads in Emerging Markets and Developing Economies[1]

Jorge E. Carrera[2], Gabriel Montes-Rojas[3] and Fernando Toledo[4]

November 2021

## Abstract

We study the diffusion of shocks in the global financial cycle and global liquidity conditions to emerging and developing economies. We show that the classification according to their external trade patterns (as commodities' net exporters or net importers) allows to evaluate the relative importance of international monetary spillovers and their impact on the domestic financial cycle volatility —i.e., the coefficient of variation of financial spreads and risks. Given the relative importance of commodity trade in the economic structure of these countries, our study reveals that the sign and size of the trade balance of commodity goods are key parameters to rationalize the impact of global financial and liquidity conditions. Hence, the sign and volume of commodity external trade will define the effect on countries' financial spreads. We implement a two-equation dynamic panel data model for 33 countries during 1999:Q1-2020:Q4 that identifies the effect of global conditions on the countries' commodities terms of trade and financial spreads, first in a direct way, and then by a feedback mechanism by which the terms of trade have an asymmetric additional influence on spreads.

[1] We would like to thank Mariquena Solla for her outstanding research work assistance. The usual disclaimer applies.
[2] CONICET, Universidad Nacional de La Plata and Banco Central de la República Argentina.
[3] Universidad de Buenos Aires and Instituto Interdisciplinario de Economía Política de Buenos Aires (IIEP-BAIRES-UBA-CONICET).
[4] Universidad Nacional de La Plata and Banco Central de la República Argentina.

## 1. Introduction

There has been increasing interest in the analysis of Global Financial Crisis (GFC) and its effects on developing countries. Monetary spillovers from the GFC are reportedly transmitted in most cases through trade and financial channels. However, some preliminary evidence shows that the impact of the GFC is not homogeneous across Emerging Markets and Developing Economies (EMDEs).

This means that the policy reaction in these economies should be carefully calibrated. In this regard, it is reasonable to wonder if clear mechanisms may explain the reason why the impact of the GFC differs across EMDEs that *prima facie* look similar. We show that the relative weight of natural resources in factor endowments—particularly through the interaction between the productive structure and external trade issues related to their configuration as net exporters or net importers of commodities—is key to understanding how the diffusion of GFC and global liquidity conditions affect financial spreads (FS) in EMDEs.

The literature distinguishes two possible strategies to proxy the GFC (Scheubel et al., 2019; Colacito et al., 2018): (1) its direct measurement from the VIX indicator[5]; and (2) its indirect estimation through dynamic factor models, the GFC (Common Global Factor, CGF) (Miranda-Agrippino and Rey, 2020). An easing (tighter) monetary policy links to a decline (rise) in the VIX. The opposite happens when we examine the behavior of the CGF. In the present paper, we use these two alternative proxies to quantify the GFC.

In the first round of analysis, we note that the GFC affects homogeneously both the commodity price index and domestic financial spreads, regardless of the productive structure and external trade patterns of the EMDEs. Figures 1 and 2 show that the correlations between the VIX and CGF and the commodity price index are negative (positive for CGF) at the usual significance levels.

The correlations between the VIX and FS also display a positive sign for net commodity exporter and net importer countries (see Figure 3).[6] This homogeneous relationship is less clear when we examine the correlations between the CGF and domestic FS (as measured by EMBIG).

---

[5] VIX stands for the Chicago Board Options Exchange Volatility Index. It is a measure used to track volatility on the S&P 500 index.

[6] To define net exporters and net importers commodities dependent EMDEs, we use the criteria employed in Bastourre et al. (2015). We identify net exporters as Group 1 —those that show a zero or positive commodity trade balance—, and net importers as Group 2 —those which have a negative commodity trade balance.

However, we see that the higher proportion of net commodity exporters and importers falls in the negative region (see Figure 4), with the exception of India.

We have synthesized the external trade patterns of each EMDE in one recent measure: The CTOT index for each country (Gruss and Kebhaj, 2019). This indicator allows us to determine how changes in the CTOT affect the EMDEs in different ways, according to the share of commodities in the exports and imports of each economy.

Thus, we find out —in a second round of GFC diffusion— that the correlation between CTOT and FS is heterogeneous and depends on the productive structure and external trade patterns through the specific CTOT variable: In net exporters (net importers), there is a negative (positive) association between the dynamics of (both) CTOT and spreads (see Figures 5 and 6).

The propagation of GFC to EMDEs through the CTOT creates an asymmetric result on the perception of financial risk and solvency, approximated by EMDEs' financial spreads. These second round effects of GFC amplify (dampen) the domestic financial cycle volatility in net exporters (net importers) economies.

We contribute to the literature in several ways. First, we expand the empirical analysis of these second round effects of GFC, CTOT and FS to disentangle the signs of these correlations according to productive structure and external trade patterns. Second, we estimate a quarterly dynamic panel model for 33 EMDEs during 1999:Q1-2020:Q4 to examine the relationship between CTOT and FS for net exporters and net importers commodities EMDEs. We empirically control the incidence of current account balance and trade openness (alternative control variables confirm the robustness of our model). Finally, we consider some alternative measures of GFC —VIX and CGF— and two additional proxies to capture changes in global liquidity conditions —Federal Funds Rate (FFR); and Broad Effective Nominal Exchange Rate for US (NERUS).

The structure of the paper is as follows. In Section 2, we review the literature on the relationships among GFC, CTOT and FS in net exporters and net importers commodities dependent EMDEs. In Section 3, we describe the database. In Section 4, we introduce our econometric strategy. In Section 5, we show the main empirical results. In Section 6, we offer some final remarks.

**2. Theoretical framework**

Monetary policy shocks in the financial centers are one of the main determinants of GFC. The monetary policy international spillovers from financial centers to EMDEs are transmitted through different channels like the commercial and the financial ones.

A fall in US monetary policy interest rate stimulates greater appetite for risk of global investors —i.e., a fall in VIX—, who decide to reallocate their financial funds mostly to developing economies. Thus, lower policy interest rates in the US trigger nominal appreciation in small open economies driven by portfolio flows and cross-border lending (Bonizzi and Kaltenbrunner, 2021; Yilmaz and Godin, 2020). These external funds are reverted when the FED announces an increase in its nominal interest rate policy —the FFR.[7] In such a case, we observe increases in global risk aversion that stimulates flight-to-quality behavior and higher exchange rate market pressure in EDMEs, even though these economies show solid macroeconomic fundamentals (Kohler, 2021; Botta, 2021).

There is a considerable pass-through from changes in global financial conditions —such as VIX or CGF— towards EMDEs economies; particularly on exchange rates, asset prices, risk premium, and credit growth (Miranda-Agrippino and Rey, 2021; and Jordà et al., 2018). These outcomes are consistent with the distinction between push and pull factors: the former concerning factors unrelated to the conditions of the recipients, and the latter referring to the variables of the recipient economy (Bruno and Shin, 2017; Aizenman et al., 2016; Calvo et al., 1996). In any case, the impact of these international financial cycles depends on specific structural characteristics of EMDEs (Cimoli et al., 2017).

In the last few decades, empirical evidence has shown that push factors —particularly the monetary stance of US— have been a major driver of ebb and flow capital movements to EMDEs (Chari et al., 2020; Rey, 2013). For instance, Aidar and Braga (2020) examine the extent to which push factors linked to global liquidity play an important role —compared to country-specific factors— in changes in the risk premium for a set of developing economies during the period 1999-2019. These authors find evidence that the common factors behind the set of country-risk premiums can be explained by financial variables, namely the US interest rate and the VIX.

Shocks on commodity prices are also important drivers of cyclical fluctuations in net exporters commodities dependent EMDEs (Roch, 2019). The impact of Terms of Trade (TOT) shocks on business cycle fluctuations depends on economies pattern of production and international trade. For example, Kohn et al. (2021) suggest that emerging economies are more vulnerable to TOT shocks given that they run significant sectoral trade imbalances, with large trade surpluses in commodities and large deficits in manufactures. Furthermore, according to Drechsel and Tenreyro (2018), commodity prices and FS show a negative correlation in net exporters commodities dependent EMDEs.

---

[7] The EMDEs' exposure to GFC increases the risk of financial crisis and raises the specter of constraints on policy autonomy (Grabel, 2019).

Last, but not least, the analysis of super cycles of commodity prices highlights the relevance of the Prebisch-Singer' declining trend TOT hypothesis and points out its relationships with changes in international capital flows. These contributions have almost been focused on net exporter countries and generally conclude that commodity prices and capital inflows to EMDEs are related to higher global liquidity conditions in US (Reinhart et al., 2020; Moreno et al., 2014; Erten and Ocampo, 2012).

The literature on this topic has focused on the impact of GFC and global liquidity conditions on EMDEs, particularly in net exporter countries. Nevertheless, there is no comprehensive treatment that shows whether production and external trade structures concerning commodities have any relevance in determining how they are processed domestically in terms of the financial effects related to GFC and global liquidity conditions.

The present paper shows that the classification of EMDEs according to their external trade patterns allows to evaluate the relative importance of international monetary spillovers and their impact on the domestic financial cycle volatility —i.e., the coefficient of variation of financial spreads and risks.[8] Given the relative importance of commodity trade in the economic structure of these countries, our study reveals that the sign and size of the trade balance of commodity goods are key parameters to rationalize the impact of GFC and global liquidity conditions. Hence, the sign and volume of commodity external trade will define the extent of changes in GFC and global liquidity conditions on countries' financial spreads.

To understand these asymmetric effects on financial risks in detail, it is important to consider both the first round direct effect of GFC and global liquidity conditions on financial spreads, and their incidence on commodities prices, on the one hand; and the secondary influence of GFC and global liquidity conditions over spreads, on the other. Thus, GFC and global liquidity conditions influence developing countries' financial risk in two ways: on a direct basis; and rather indirectly through commodities prices.

---

[8] According to our estimations, in net commodity exporters the volatility of financial spreads —the coefficient of variation of the EMBIG for the completed period— is 0.69. For net commodity importers, its value equals 0.49.

In the case of net exporters, the boom phase of GFC and improvements in global liquidity conditions led to a reduction in FFR and brought about a simultaneous increase in commodities prices. The first round effects cause a homogeneous —and relatively quick— fall in countries' financial risk. Hence, GFC and global liquidity conditions expand in a positive and parallel way by both channels: higher commodities prices gradually improve the current account, and lower FS also do so through the financial account in each of these EMDEs. In a second round, the increase in commodity prices and the improvement in the value of exports create positive feedback on spreads. This is explained by a gradual rise in commodities' revenues that consolidate the perception of greater solvency, leading to further enhanced financial conditions. So, in net exporters EMDEs, we find a negative correlation between commodity prices and financial spreads.

In net importers, positive shocks to GFC and global liquidity conditions enable the access to external financing due to lower international interest rates, thus reducing financial spreads at the first round. Moreover, the higher commodity prices gradually deteriorate the trade balance in these EMDEs. Thus, the increase in the cost of the imported commodities progressively deteriorates the current account in a second round, which worsens the perception of solvency and gives negative feedback to financial spreads.

3. **Data description**

We employ a quarterly panel database including 33 EMDEs for the period 1991:Q1-2020:Q4. A detailed description of the countries, variables and their sources is in the Appendix.

We use two variables to quantify GFC: the log of VIX, and the CGF computed by Miranda-Agrippino and Rey (2020). To approximate the changes in global liquidity we employ the US Effective FFR and the NERUS.

To define net exporters and net importers, we apply the criteria described by Bastourre et al. (2015). We classify net exporters as Group 1 —those that show a zero or positive commodity trade balance—, and net importers as Group 2 —those who have a negative commodity trade balance— on the basis of the COMTRADE annual data about external trade flows of 48 commodities. We implement the annual ratio between net exports of commodities and total trade flows of commodities for each EMDEs during each year to classify these economies according to this criterion. See the Appendix for the countries classification.

To account for countries' financial spreads, we use the EMBIG—the JP Morgan EMBI Global Sovereign Spread index blended spread— extracted from Bloomberg and The Global Economic Monitor.

Finally, to measure the commodity price index, we use the Data on Primary Commodity Prices based on the IMF's Primary Commodity Price System. In addition, we follow the methodology of Gruss and Kebhaj (2019) to identify commodity prices using the IMF CTOT database. We adopt two indicators to measure CTOT; CTOT1: Commodity Net Export Price Index, Individual Commodities Weighted by Ratio of Net Exports to Total Commodity Trade; and CTOT2: Commodity Net Export Price Index, Individual Commodities Weighted by Ratio of Net Exports to GDP. Note that these two variables are country-specific and time-varying according to the bundle of net export commodities each year.

## 4. Econometric strategy

We consider a quarterly panel data model with two country-level endogenous variables $(p_{it}, y_{it})$ where $p$ denotes the country-specific commodity price index and $y$ stands for the country-specific financial risk measure (EMBIG), $i = 1, ..., N$ indexes countries and t= 1, ..., $T$ time. For our particular case $N$=33 and $T$ varies between 5 and 39 (with an average of 33), thus resulting in an unbalanced panel.

The main interest corresponds to a change in GFC and global liquidity conditions, which will be studied by different variables denoted by $r$.

The system of equations we want to estimate is as follows:

$$p_{it} = \sum_{j=1}^{L} a_j^p p_{it-j} + \sum_{j=0}^{L}\left(a_j^r r_{t-j} + a_j^{rx} r_{t-j} \mathbf{1}[exporter_i]\right) + AX_{it} + \mu_i + \varepsilon_{it}, \quad (1)$$

$$y_{it} = \sum_{j=1}^{L} b_j^y y_{it-j} + \sum_{j=0}^{L}\left(b_j^r r_{t-j} + b_j^{rx} r_{t-j} \mathbf{1}[exporter_i]\right) + \sum_{j=0}^{L}\left(b_j^p p_{t-j} + b_j^{px} p_{t-j} \mathbf{1}[exporter_i]\right) + BX_{it} + \lambda_i + \varepsilon_{it}. \quad (2)$$

The first equation corresponds to the commodity price dynamics, which is affected by its own lag structure, and a contemporaneous and lagged effect on the $r$ variable. The second equation models the EMBIG dynamics. The model has been developed by assuming that $y$ has no effect on $p$, while $p$ has a contemporaneous (and lagged) effect on $y$. A distinctive feature of the model is that $r$ is assumed to have a potential different effect on $p$ and $y$ depending on whether the country is classified as net exporter or net importer. This is modelled using the interaction of $r$ with a dummy that identifies the country's classification. Moreover, commodity prices may affect the country financial conditions differently depending on its net exporter status as well. The model considers country specific fixed-effects, $\mu$ and $\lambda$ for the commodity prices and financial spreads equations, respectively.

We consider two specifications that change the sample size according to data availability. First, we draw on a model without control variables $X$, for which we have a sample of 33 countries. Second, we use a common set of control variables $X$ that has country current account balance as GDP and trade openness (also divided by GDP). For this case the sample size reduces to 19 countries due to the availability of quarterly data for the period of analysis. Although not reported, we also consider different control variables with the same results. In particular, we add international reserves and financial market conditions in the US (S&P 500). Results are available from the authors upon request. The common number of lags for both models is $L$=2.

For the $r$ variable we consider two approaches with four different variables. First, we use two variables related to GFC, the log of VIX, and the CGF—the latter being interpreted with the reverse sign concerning the log of VIX. Second, we contemplate two variables related to global liquidity conditions, FFR and NERUS. FFR can be interpreted as a measure of US monetary policy stance and indicates that a positive shock corresponds to tightening the global liquidity or financial conditions. Furthermore, when FFR increases, NERUS goes down, which means a nominal appreciation of the US dollar. The empirical model also assumes that the EMDEs countries' terms of trade and financial spreads do not affect the global financial and liquidity conditions.

In order to account for potential dynamic panel bias, the two equations are estimated by the System GMM method of Arellano and Bond (1991) and Blundell and Bond (1998). In this set-up, lagged values of the dependent variable are used as instruments for the endogenous variables. For the first equation, $r$, $r \times \mathbf{1}[exporter]$ and $X$ are treated as exogenous. For the second equation, $p$, $p \times \mathbf{1}[exporter]$ $r$, $r \times \mathbf{1}[exporter]$ and $X$ are treated as exogenous. We use the collapse instrument option in Roodman (2009a, b) to avoid the potential effects of many instruments.[9]

---

[9] In all cases the Hansen test cannot reject the null hypothesis of validity of the instrument set and the set of assumed exogenous variables. The AR(2) test for validity of the moment conditions cannot reject the null of valid moments constructed from lagged values.

Our main interest lies in computing the impulse-response functions (IRFs) of a positive shock in $r$ given by a sample standard deviation. We compute the direct impact of this shock on the bivariate system, $r \to (p, y)$ and also the particular effect that this shock has on $y$ only through $p$, $r \to p \to y$. For the latter we set $b_j^r = b_j^{rx} = 0$, $j = 0, 1, 2$ in the IRFs computation in order to isolate the distinctive effect of commodity prices on countries' EMBIG. For the purpose of constructing confidence intervals, we resort to parametric bootstrap using 200 replications, where the estimated coefficients in each equation are randomly drawn from a multivariate normal distribution with the estimated (robust) variance-covariance matrix calculated from the GMM model. We compute the confidence intervals using 0.1 and 0.9 quantiles of the simulated distribution of the effects.

## 5. Empirical results

The results are presented in terms of IRFs. Figures 7 to 14 relate to a positive shock in log VIX, CGF, FFR and NERUS, which act as different proxies of GFC and global liquidity conditions, given by $r$. Figures 7 to 10 use the specification without additional controls, and a sample size of 33 countries as well. Figures 11 to 14 include current account balance and trade openness (in both cases as a share in GDP), with a sample size of 19 countries on account of data availability limitations. In each figure, the first row corresponds to the CTOT1 commodity price index, while the second row to CTOT2, both used for the variable $p$ in the econometric model. Column A in each figure corresponds to the effect on commodity prices ($r \to p$), separately for net exporters and net importers. Column B in each figure corresponds to the total effect on countries' EMBIG ($r \to y$), also separately for net exporters and net importers. Finally, Column C computes the effect on countries' EMBIG that comes only through the price channel effect ($r \to p \to y$) for the same two groups of countries.

### 5.1 Impact on commodities prices ($r \to p$)

Let's consider first the effect of a positive shock in $r$ on commodities' prices, CTOT1 and CTOT2 (Columns A). In general, it has a differential effect on net exporters *vis-à-vis* net importers. One positive standard deviation shock in $r$ has a negative effect on the former (reducing commodity prices for exporters) and a positive effect on the latter.

The dynamics differ by the type of shock. A positive shock in log VIX and NERUS produces a transitory reduction in both CTOT1 and CTOT2 for net exporters, but a rise for net importers. A positive shock in CGF (interpreted with the reverse sign) and FFR, however, produces a permanent effect on both CTOT variables, with a decline for net exporters and a rise for net importers.

## 5.2 Impact on countries' financial risk ($r \rightarrow y$)

The effect on the EMBIG is of the expected sign (see Columns B), implying that the bust phase of GFC (a tightening in the global financial conditions) increase countries' financial risk. The effects are, in general, larger and more persistent for net exporters than for net importers. All of the $r$ variables, except FFR, have a positive (i.e., increasing country risk) statistically significant contemporaneous and short-run effect. In the case of the log of VIX, the long-run effect is positive for both net exporters and net importers, but in the case of FFR and NERUS there is a positive impact on exporters but a negative one on importers. There is no clear long-run persistence when using CGF proxy.

Differences in the dynamic paths point out to the existence of first-order and second-order effects. First, there is a clear financial risk effect possibly reflecting the flight-to-quality pattern of capital flows. The boom phase of GFC and a tightening of global liquidity conditions clearly affect all EMDEs, independent of their international trade insertion. However, given the differential effects arising from the CTOT channel, there is an overall differentiation between net exporters and net importers. This is a second-order effect and is the main goal of this paper. In order to study this particular effect, we isolate the effect of GFC and global liquidity conditions on the EMBIG variable that comes only through the commodities prices.

## 5.3 Impact on countries' financial risk through commodities prices ($r \rightarrow p \rightarrow y$)

Finally, we study the effect of GFC and global liquidity conditions on EMBIG using only the CTOT channel (see Columns C). The results are smaller and weaker than the total effects ($r \rightarrow y$) in Column B. Nevertheless, in this case there is a positive influence on net exporters (implying that EMBIG increases) and a non-statistically significant effect on net importers. These results highlight that there is a specific channel arising from commodity prices, which also differs in terms of each country' net commodity trade position.

## 6. Final remarks

In the present paper, we show that the relative weight of natural resources in factor endowments —a key parameter that defines the external trade pattern of each country— is key to understanding how the diffusion of GFC and global liquidity conditions affect financial volatility in EMDEs, particularly through the interaction between the productive structure and external trade issues related to their configuration as net exporters or net importers of commodities. This classification of EMDEs is a key variable to determine the effect of the GFC in EMDEs. We develop a simple empirical model that shows the impact of the GFC on the CTOT and financial spreads, first in a direct way, and then through a feedback mechanism: The CTOT influence on the financial spreads.

In general, the effect of a positive shock in $r$ on CTOT has a differential effect on net exporters *vis-à-vis* net importers. One positive standard deviation shock in $r$ has a negative effect on the former (reducing the price of commodities for exporters) and a positive effect for importers. We also find that the bust phase of GFC and tightening of global liquidity conditions increase countries' financial risk. The effects are in general larger and more persistent for net exporters than for net importers. Therefore, countries that are intensive in the external trade of commodities have a wider business and financial cycle. Inside this group, commodity net exporter countries display the most volatile fluctuations.

Last, but not least, we study the effect of the GFC and global liquidity conditions on EMBIG using only the CTOT channel. The results are smaller and weaker than the total effects. Nevertheless, there is a positive effect on net exporters (implying that EMBIG increases) and a non-statistically significant effect on net importers.

Our findings are robust to the use of different proxies for the GFC, to two alternative CTOT indices and to the inclusion of two control variables (current account balance and trade openness), which reduces the estimation sample from 33 to 19 EMDEs due to data availability limitations.

We encourage further research to analyze whether this behavior is extensible to EMDEs that are more or less intensive in trade of manufactured goods. The effects are expected to be softer in light of the documented lower volatility of manufactured goods on account of different pricing mechanisms.

**Appendix**

Net exporters and net importers EMDEs

| Country | Region | Definition | Group |
|---|---|---|---|
| Argentina | South America | Net commodity exporter | 1 |
| Brazil | Eastern Europe | Net commodity exporter | 1 |
| Bulgaria | South America | Net commodity importer | 2 |
| Chile | South America | Net commodity exporter | 1 |
| China | Asia Oriental | Net commodity importer | 2 |
| Colombia | West Africa | Net commodity exporter | 1 |
| Côte d'Ivoire | South America | Net commodity exporter | 1 |
| Croatia | Caribbean | Net commodity importer | 2 |
| Dominican Republic | South America | Net commodity importer | 2 |
| Ecuador | Northern Africa | Net commodity exporter | 1 |
| Egypt | Southern Europe | Net commodity importer | 2 |
| El Salvador | Eastern Europe | Net commodity importer | 2 |
| Hungary | South-Eastern Asia | Net commodity importer | 2 |
| India | Southern Asia | Net commodity importer | 2 |
| Indonesia | Asia Oriental | Net commodity exporter | 1 |
| Lebanon | Northern Africa | Net commodity importer | 2 |
| Malaysia | Central America | Net commodity exporter | 1 |
| Mexico | South-Eastern Asia | Net commodity exporter | 1 |
| Morocco | West Africa | Net commodity importer | 2 |
| Nigeria | Southern Asia | Net commodity exporter | 1 |

| Country | Region | Definition | |
|---|---|---|---|
| Pakistan | Central America | Net commodity importer | 2 |
| Panama | South America | Net commodity importer | 2 |
| Peru | South-Eastern Asia | Net commodity exporter | 1 |
| Philippines | Eastern Europe | Net commodity importer | 2 |
| Poland | Eastern Europe | Net commodity importer | 2 |
| Russian Federation | Central America | Net commodity exporter | 1 |
| South Africa | South-Eastern Asia | Net commodity exporter | 1 |
| Tunisia | Western Asia | Net commodity importer | 2 |
| Turkey | Western Asia | Net commodity importer | 2 |
| Ukraine | Eastern Europe | Net commodity importer | 2 |
| Uruguay | South America | Net commodity exporter | 1 |
| Venezuela | South-Eastern Asia | Net commodity exporter | 1 |
| Vietnam | Southern Africa | Net commodity exporter | 1 |

Most important net exporters and net importers EMDEs

| Country | Region | Ratio between net exports of commodities and total net exports of commodities (COMTRADE data) | Definiton |
|---|---|---|---|
| Argentina | Latin America & Caribbean | 67% | Net commodity exporter |
| Brazil | Latin America & Caribbean | 46% | Net commodity exporter |
| Colombia | Latin America & Caribbean | 55% | Net commodity exporter |
| Russian Federation | Europe & Central Asia | 84% | Net commodity exporter |
| Turkey | Europe & Central Asia | -56% | Net commodity importer |

| | | | | |
|---|---|---|---|---|
| India | South Asia | -52% | | Net commodity importer |
| Morocco | Middle East & North Africa | -75% | | Net commodity importer |
| China | East Asia & Pacific | -70% | | Net commodity importer |

Variables, time frequency and data sources

| Variable | Definition | Time frequency | Source |
|---|---|---|---|
| VIX | Chicago Board Options Exchange Volatility Index | 1999:Q1-2020:Q4 | FRED St. Louis |
| CGF | Common Global Factor extracted from a collection of 858 asset price series spread over Asia Pacific, Australia, Europe, Latin America, North America, commodity and corporate samples | 1999:Q1-2018Q4 | Miranda-Agrippino and Rey (2020) |
| Effective Federal Funds Rate | Interest rate banks charge each other for overnight loans to meet their reserve requirements (in percentage) | 1999:Q1-2020:Q4 | FRED St. Louis |
| US nominal exchange rate | Broad Effective Nominal Exchange Rate for United States | 1999:Q1-2020:Q4 | FRED St. Louis |
| Commodity price index | Commodity price index, 2016 = 100, includes both fuel and non-fuel price indices | 1999:Q1-2020:Q4 | Data on Primary Commodity Prices based on the IMF's Primary Commodity Price System |
| CTOT | Commodity Net Export Price Index, Individual Commodities Weighted by Ratio of Net Exports to Total Commodity Trade (CTOT1) | 1999:Q1-2020:Q4 | IMF (CTOT database) |

| | Commodity Net Export Price Index, Individual Commodities Weighted by Ratio of Net Exports to GDP (CTOT2) | | |
|---|---|---|---|
| FS | JP Morgan EMBI Global Sovereign Spread index blended spread (in percentage) | 1999:Q1-2020:Q4 | Bloomberg The Global Economic Monitor (The World Bank) |
| Volatility of the domestic FS | Coefficient of variation of the domestic FS | 1999:Q1-2020:Q4 | Own calculations based on Bloomberg and The Global Economic Monitor (The World Bank) |
| Current account balance/GDP | Current Account, Goods and Services, Net, US Dollars/Gross Domestic Product, Nominal, Unadjusted, US Dollars | 1999:Q1-2020:Q4 | International Financial Statistics (IFS) |
| Trade openness/GDP | Exports of Goods and Services, Nominal, Unadjusted, US Dollars Plus Imports of Goods and Services, Nominal, Unadjusted, US Dollars/Gross Domestic Product, Nominal, Unadjusted, US Dollars | 1999q1-2020q4 | International Financial Statistics (IFS) |

Figures

**Figure 1.** Correlations between the VIX (right axis) and the commodity price index (quarterly averages)

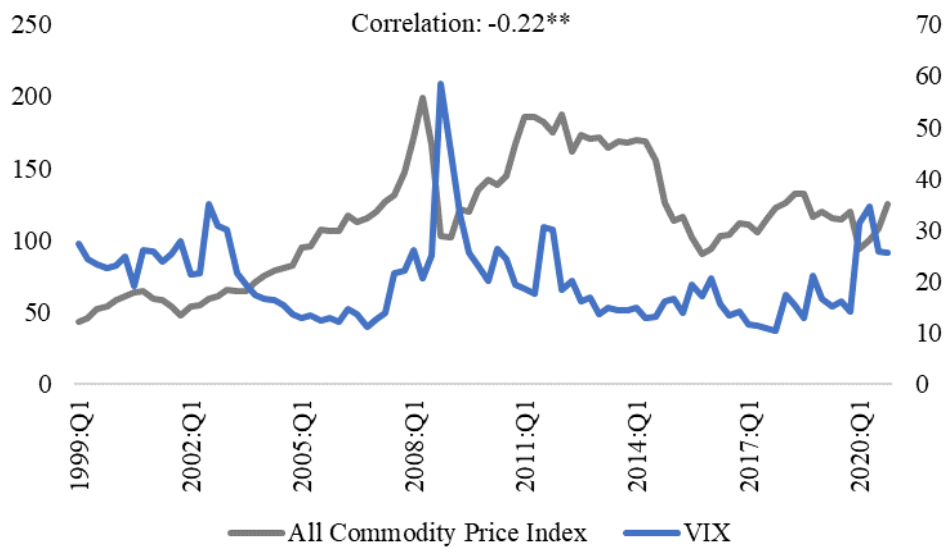

Source: FRED St. Louis and IMF. Commodity price index, 2016 = 100, includes both fuel and non-fuel price indices. ** significant at 5%.

**Figure 2.** Correlations between the CGF (right axis) and the commodity price index (quarterly averages)

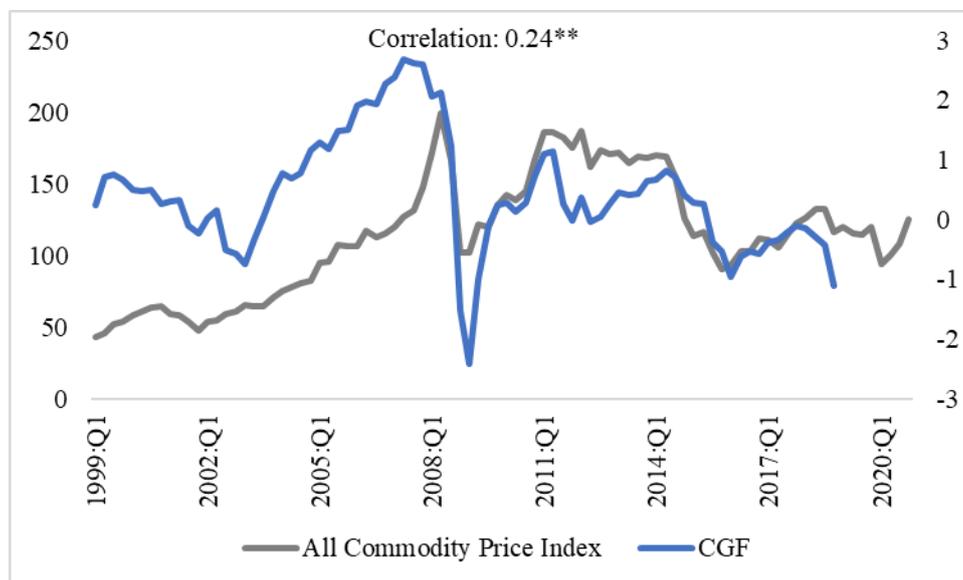

Source: FRED St. Louis and IMF. Commodity price index, 2016 = 100, includes both fuel and non-fuel price indices. ** significant at 5%.

**Figure 3.** Correlations between the VIX and the FS for net exporters and net importers commodities dependent EMDEs

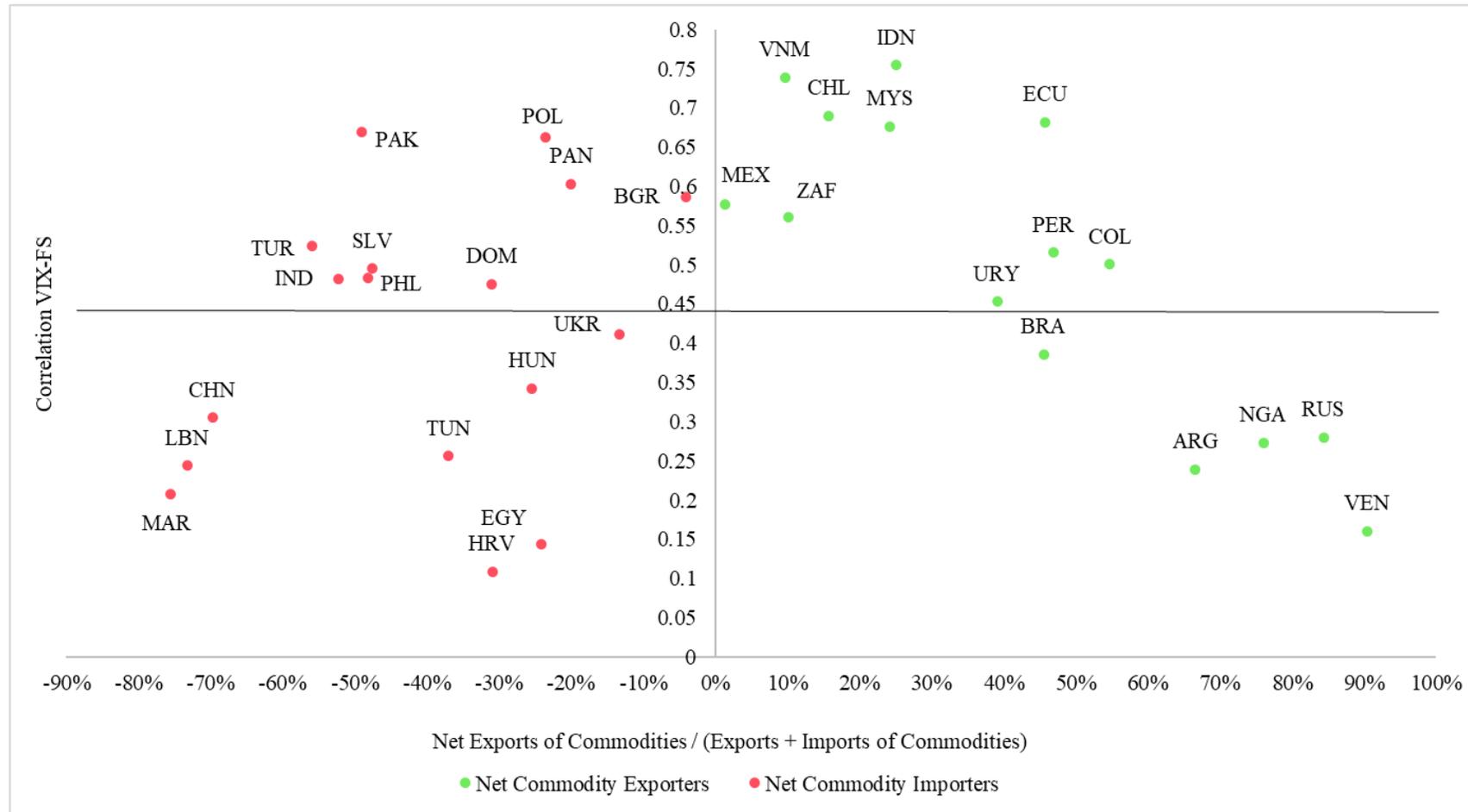

Source: FRED St. Louis, CTOT database (IMF), Bloomberg, the Global Economic Monitor (The World Bank), and COMTRADE. The solid black line shows the simple average value of the correlations between the VIX and the FS for all the EMDEs.

**Figure 4.** Correlations between the CGF and the FS for net exporters and net importers commodities dependent EMDEs

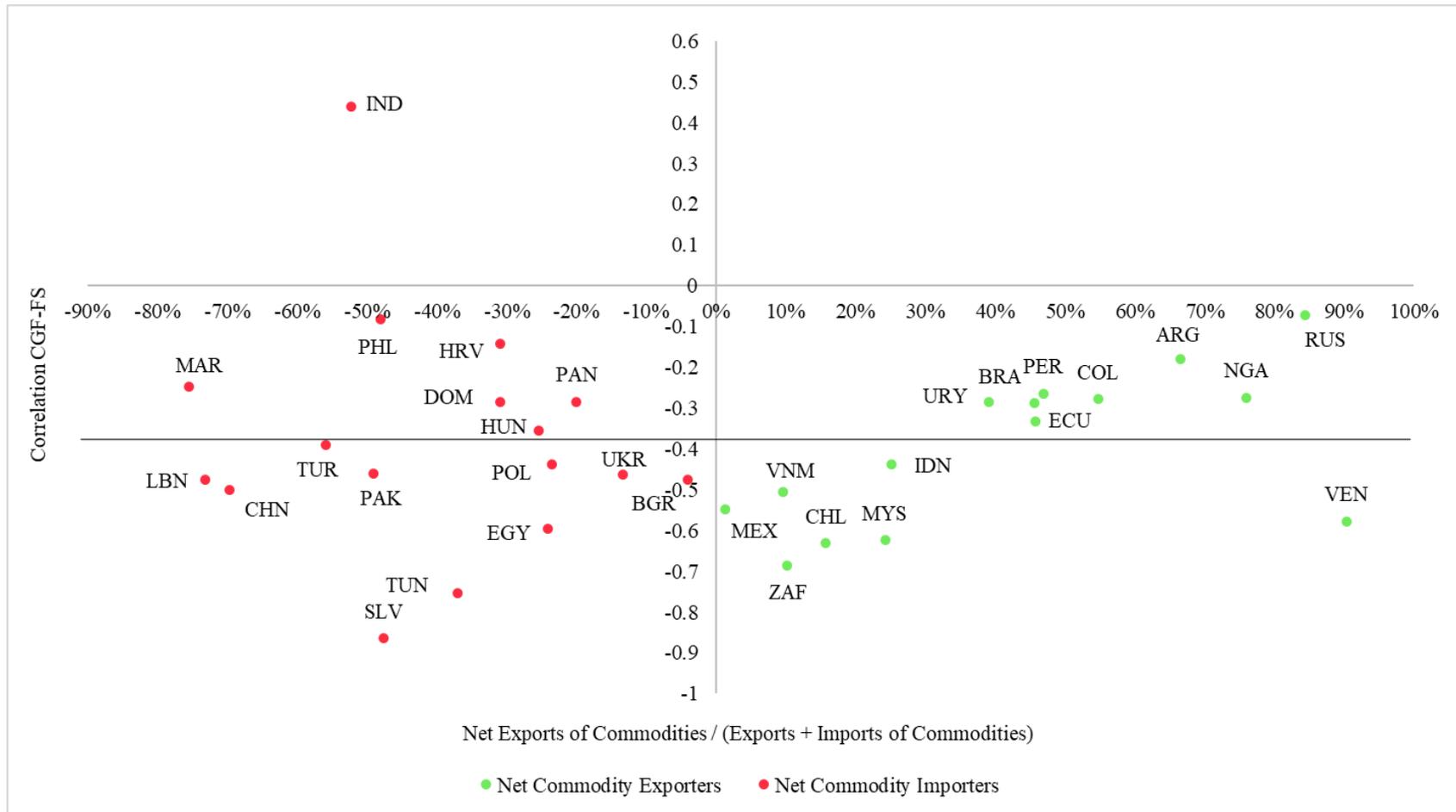

Source: FRED St. Louis, CTOT database (IMF), Bloomberg, the Global Economic Monitor (The World Bank), and COMTRADE. The solid black line shows the simple average value of the correlations between the VIX and the CGF for all the EMDEs.

**Figure 5.** CTOT (right axis) and FS (financial spreads) for net exporters commodities dependent EMDEs

(quarterly averages)

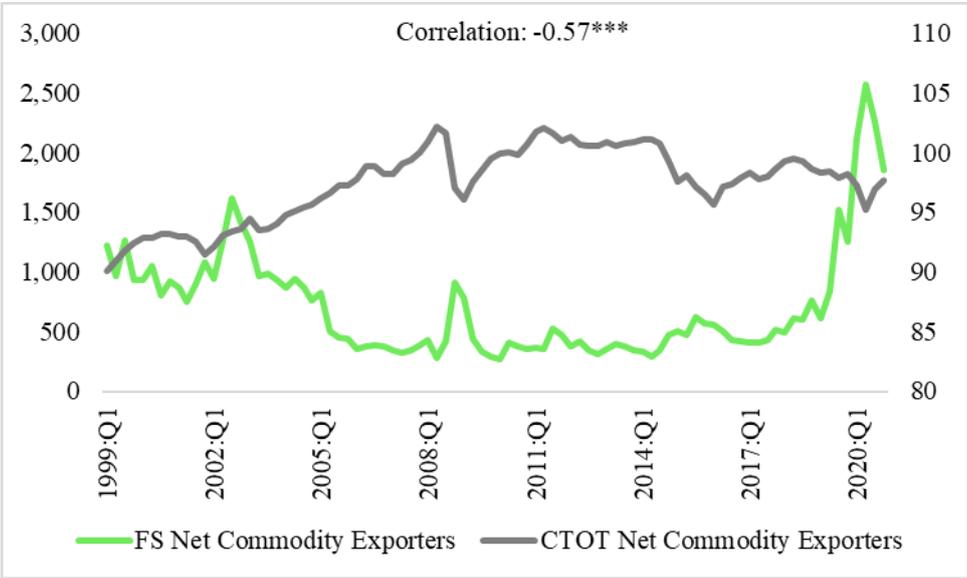

Source: CTOT database (IMF), Bloomberg, the Global Economic Monitor (The World Bank), and COMTRADE. *** significant at 1%.

**Figure 6.** CTOT (right axis) and FS (financial spreads) for net importers commodities dependent EMDEs

(quarterly averages)

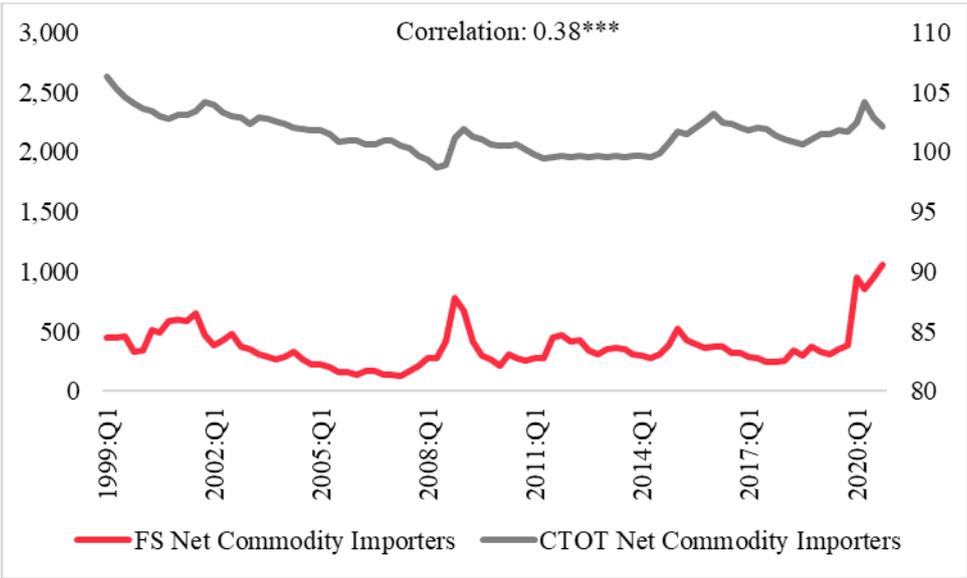

Source: CTOT database (IMF), Bloomberg, the Global Economic Monitor (The World Bank), and COMTRADE. *** significant at 1%.

**Figure 7: r: log of the VIX index**

A: $r \to p$  B: $r \to y$  C: $r \to p \to y$

**CTOT1**

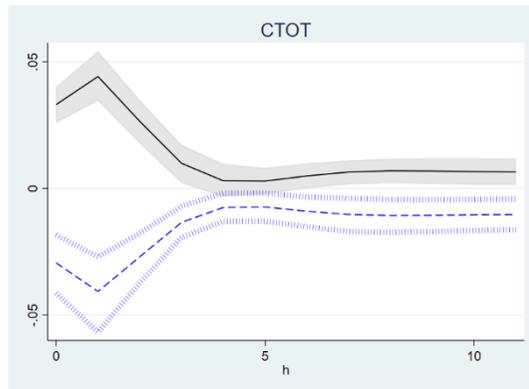 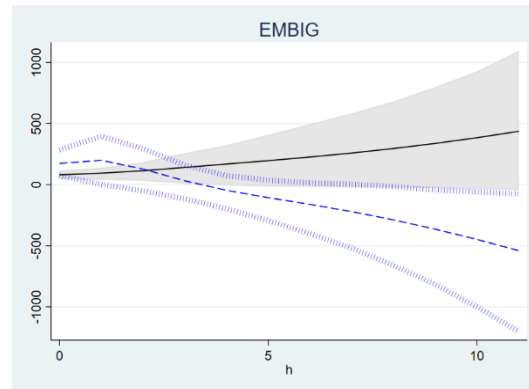 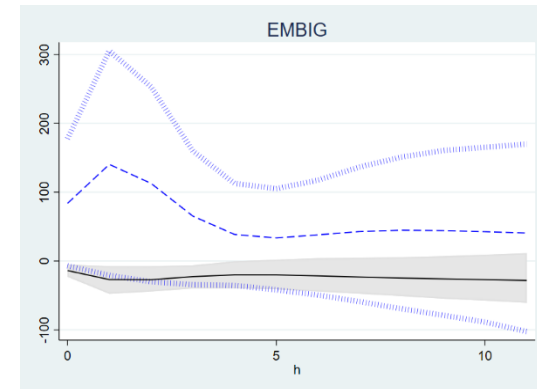

**CTOT2**

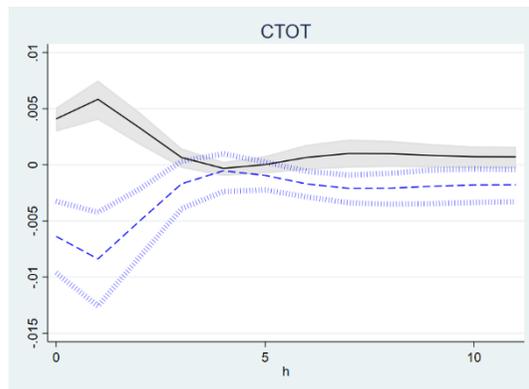 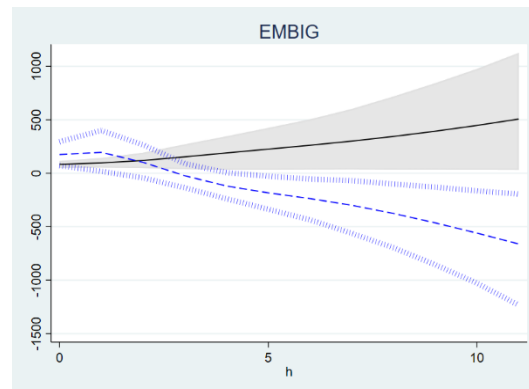 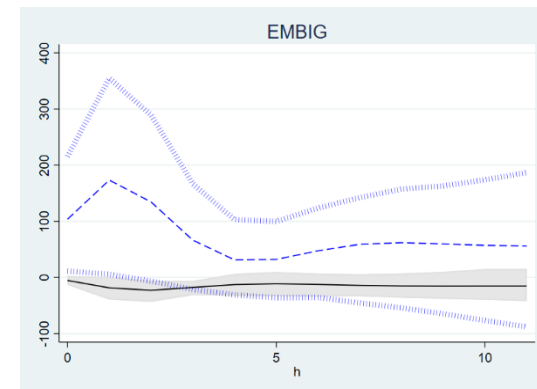

Notes: Solid lines show the IRFs for net importers and dashed lines depict the IRFs for net exporters. CTOT1: Commodity Net Export Price Index, Individual Commodities Weighted by Ratio of Net Exports to Total Commodity Trade; and CTOT2: Commodity Net Export Price Index, Individual Commodities Weighted by Ratio of Net Exports to GDP. The sample has 33 countries. The countries are Argentina, Brazil, Bulgaria, Chile, China, Colombia, Côte d'Ivoire, Croatia, Dominican Republic, Ecuador, Egypt, El Salvador, Hungary, India, Indonesia, Lebanon, Malaysia, Mexico, Morocco, Nigeria, Pakistan, Panama, Peru, Philippines, Poland, Russian Federation, South Africa, Tunisia, Turkey, Ukraine, Uruguay, Venezuela and Vietnam.

**Figure 8: r: Common Global Factor (CGF)**

| A: $r \to p$ | B: $r \to y$ | C: $r \to p \to y$ |

### CTOT1

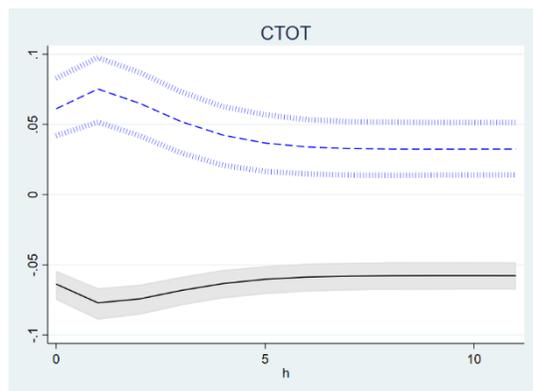
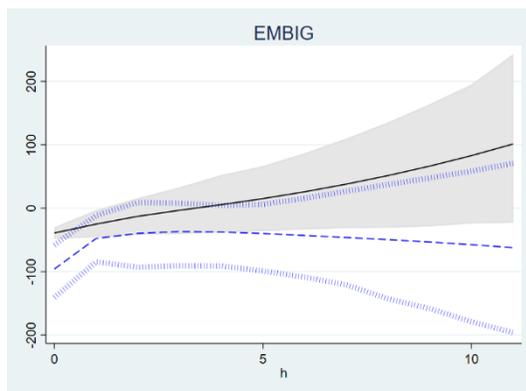
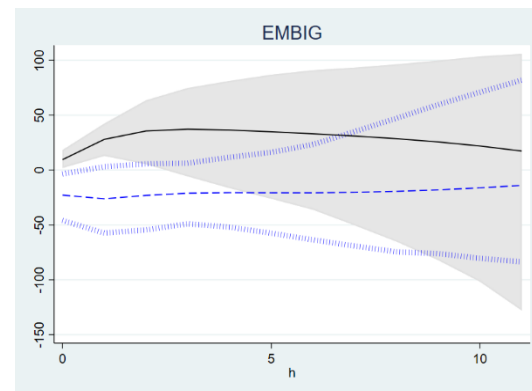

### CTOT2

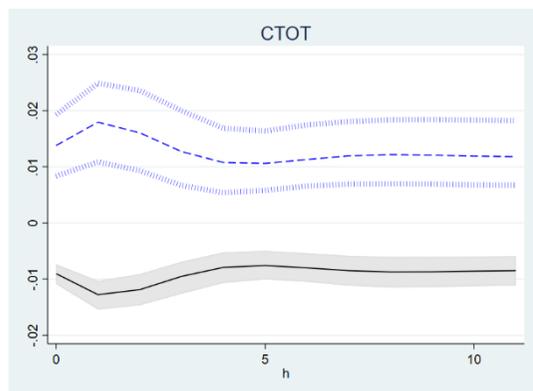
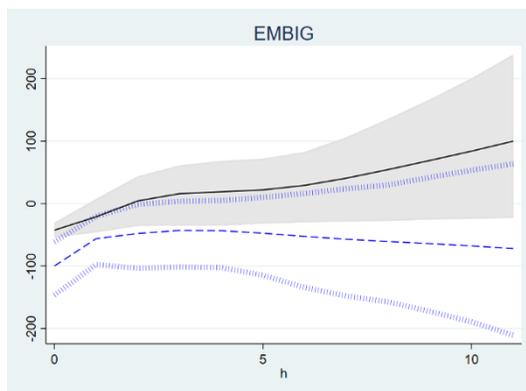
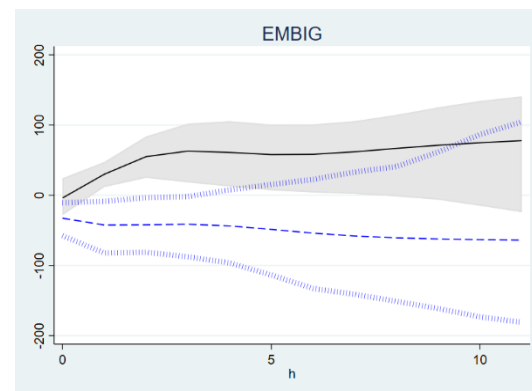

Notes: See notes to Figure 7.

# Figure 9: r: Effective Federal Funds Rate (FFR)

| A: $r \to p$ | B: $r \to y$ | C: $r \to p \to y$ |

**CTOT1**

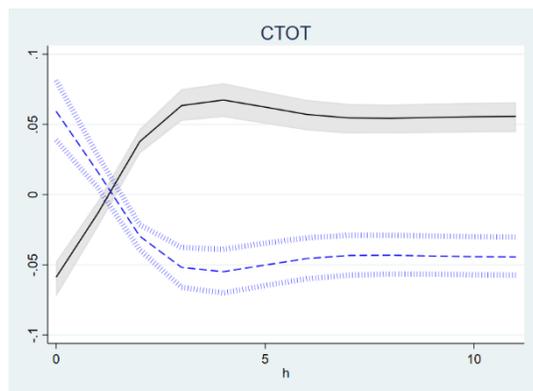 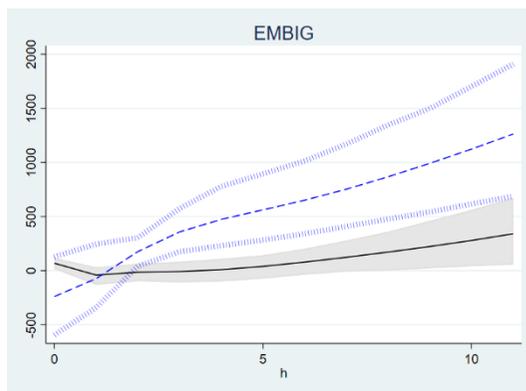 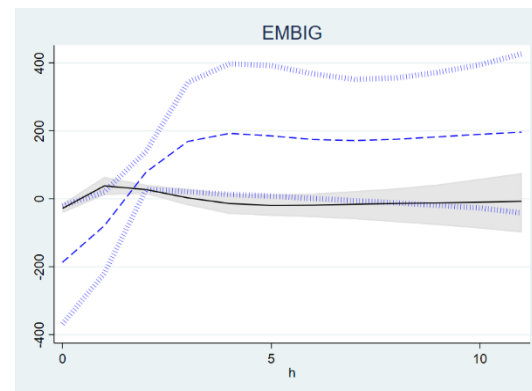

**CTOT2**

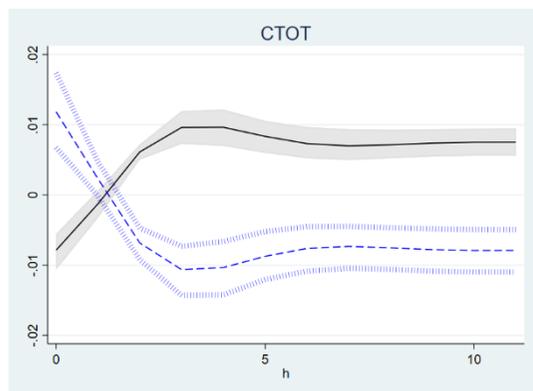 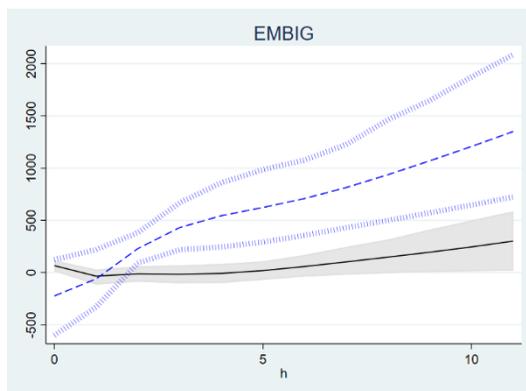 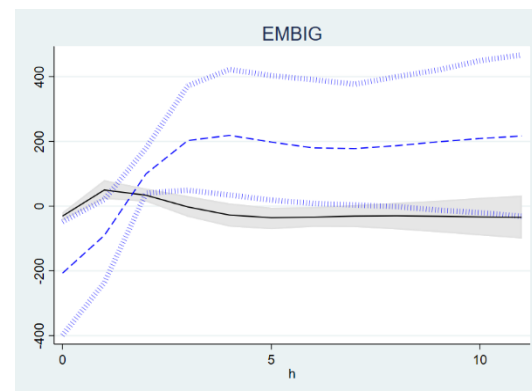

Notes: See notes to Figure 7.

**Figure 10: r: Nominal Exchange Rate for US**

A: $r \to p$  B: $r \to y$  C: $r \to p \to y$

**CTOT1**

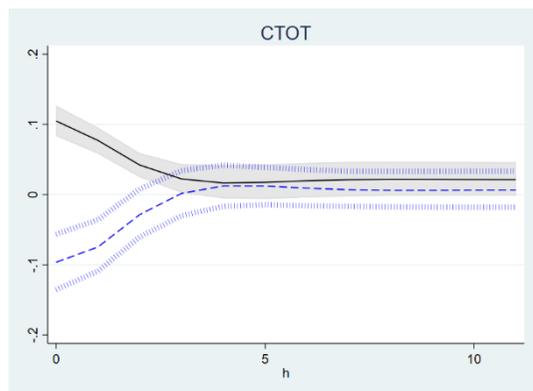 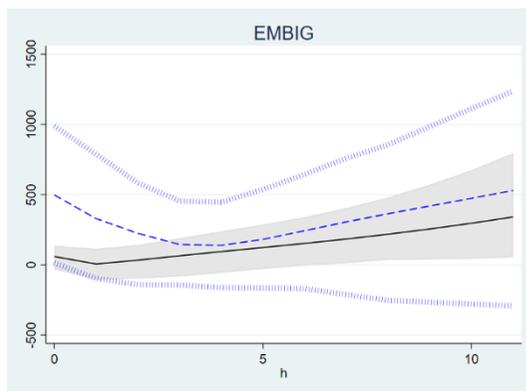 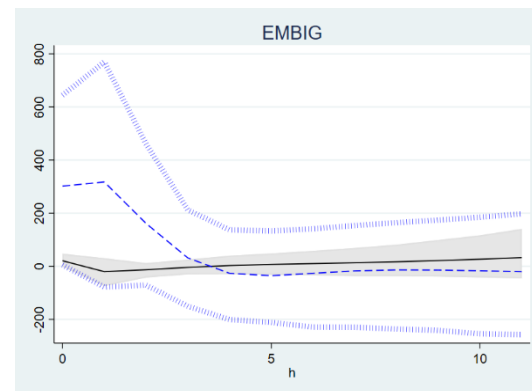

**CTOT2**

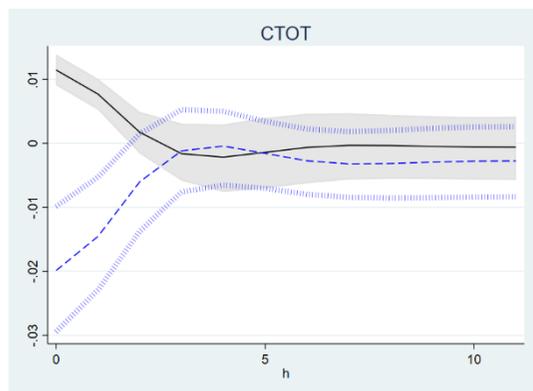 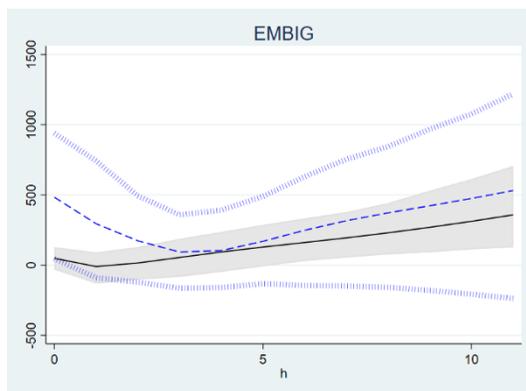 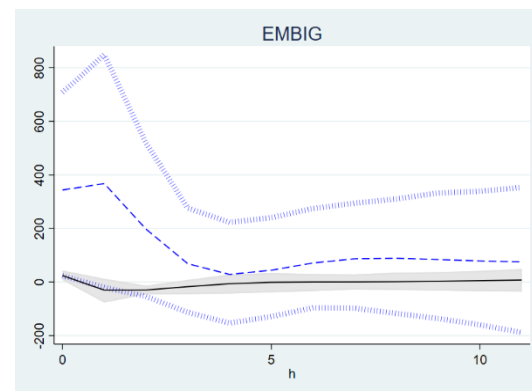

Notes: See notes to Figure 7.

**Figure 11: r: log of the VIX index**

| A: $r \to p$ | B: $r \to y$ | C: $r \to p \to y$ |

**CTOT1**

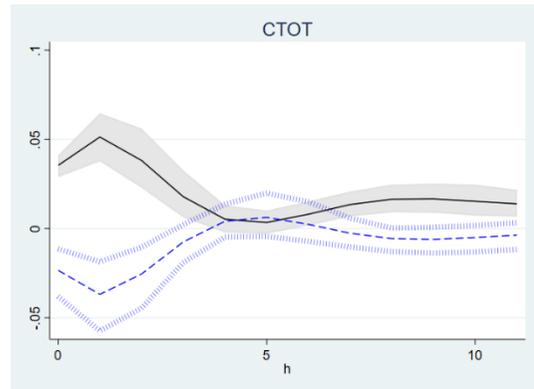 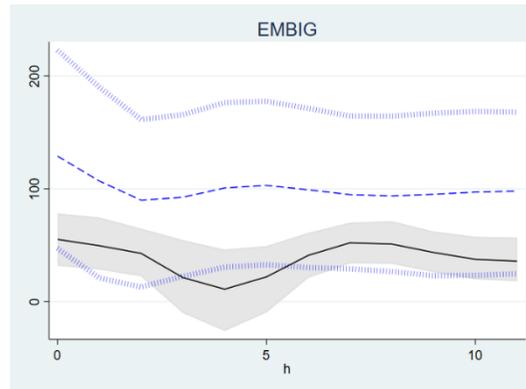 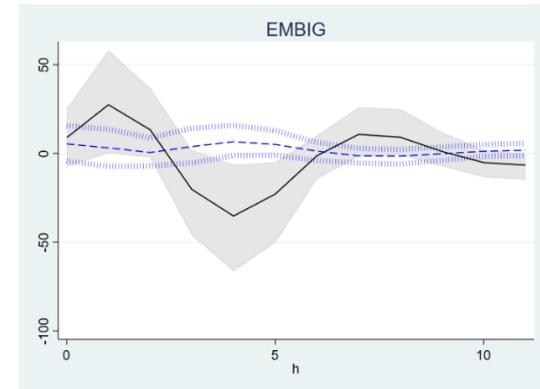

**CTOT2**

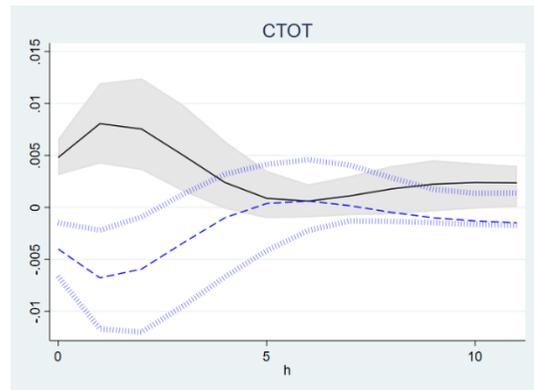 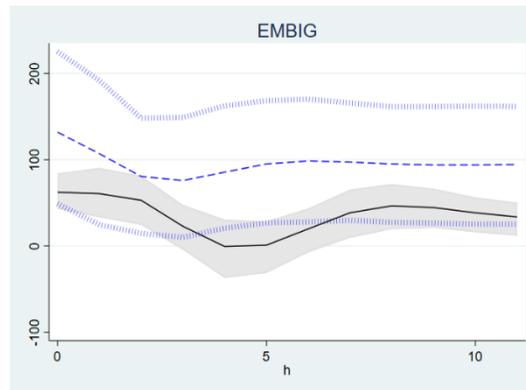 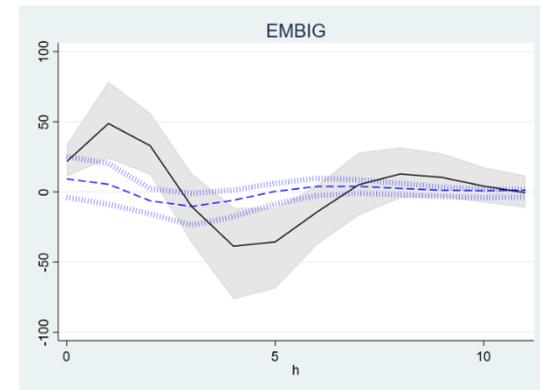

Notes: Solid lines show the IRFs for net importers and dashed lines depict the IRFs for net exporters. CTOT1: Commodity Net Export Price Index, Individual Commodities Weighted by Ratio of Net Exports to Total Commodity Trade; and CTOT2: Commodity Net Export Price Index, Individual Commodities Weighted by Ratio of Net Exports to GDP. The sample contains 19 countries and controls for current account balance and trade openness (both as percentage of GDP). The countries are Argentina, Brazil, Bulgaria, Chile, Colombia, Croatia, Ecuador, El Salvador, Hungary, India, Indonesia, Mexico, Philippines, Poland, Russian Federation, South Africa, Turkey, Ukraine and Uruguay.

**Figure 12: r: Common Global Factor (CGF)**

| A: $r \to p$ | B: $r \to y$ | C: $r \to p \to y$ |

**CTOT1**

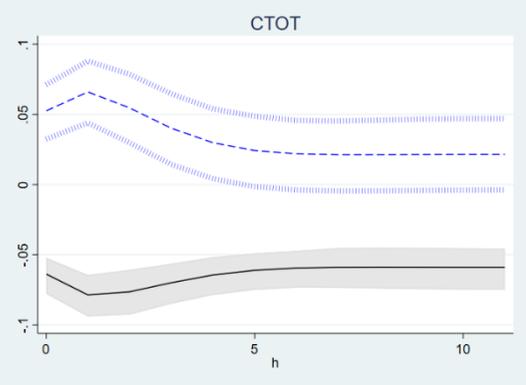 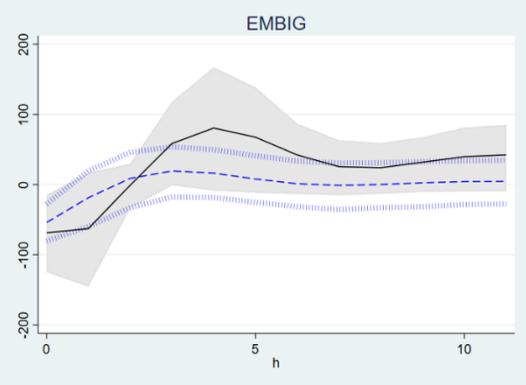 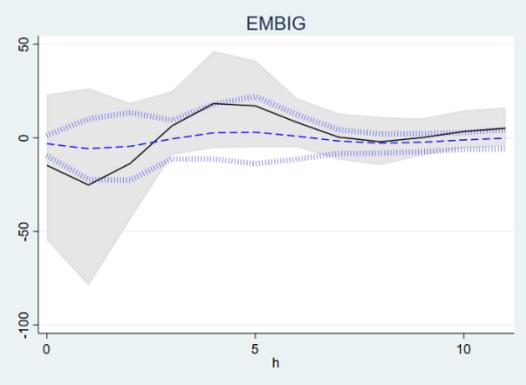

**CTOT2**

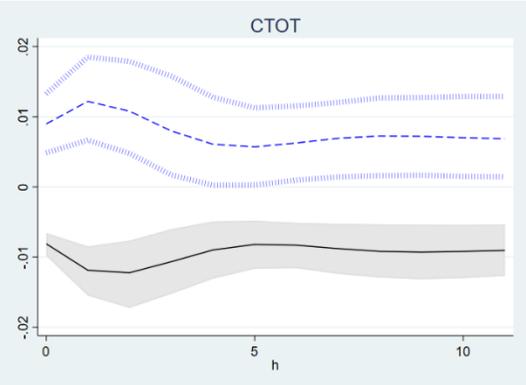 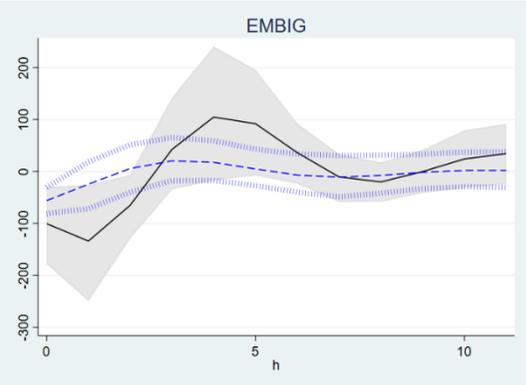 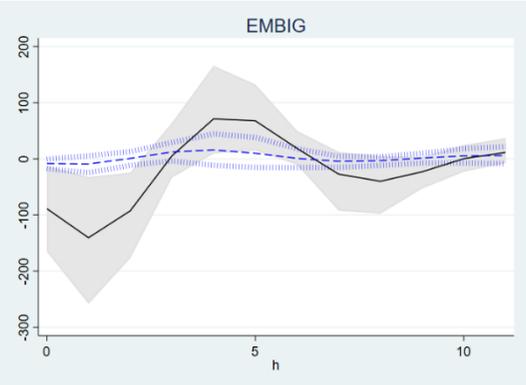

Notes: See notes to Figure 11.

**Figure 13: r: Effective Federal Funds Rate (FFR)**

A: $r \to p$          B: $r \to y$          C: $r \to p \to y$

**CTOT1**

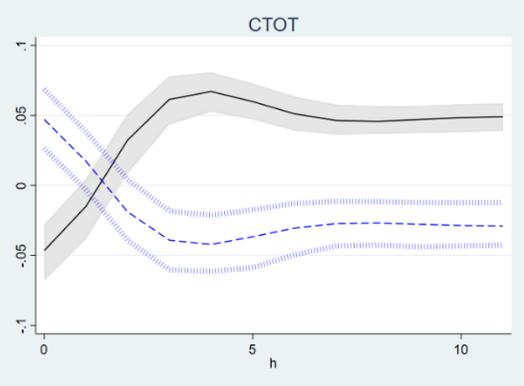 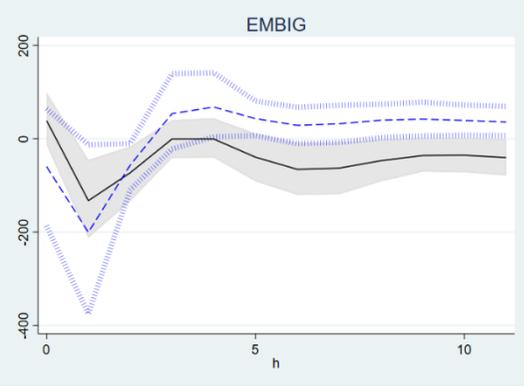 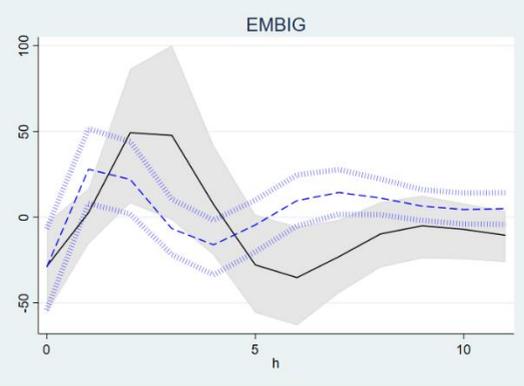

**CTOT2**

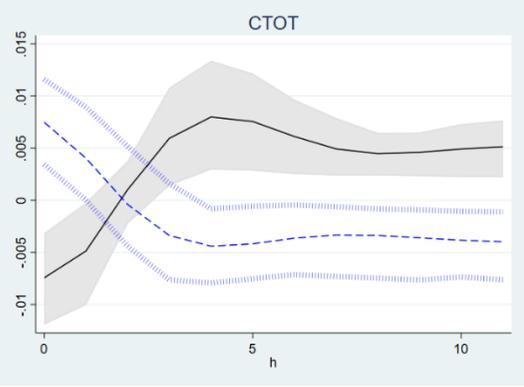 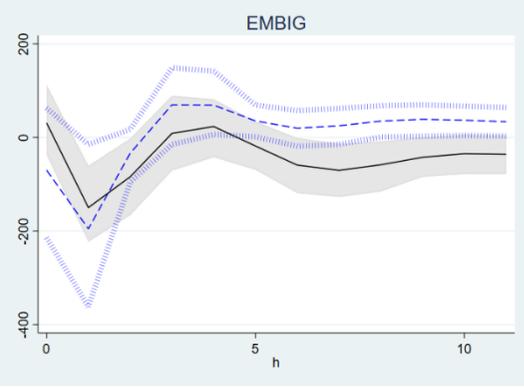 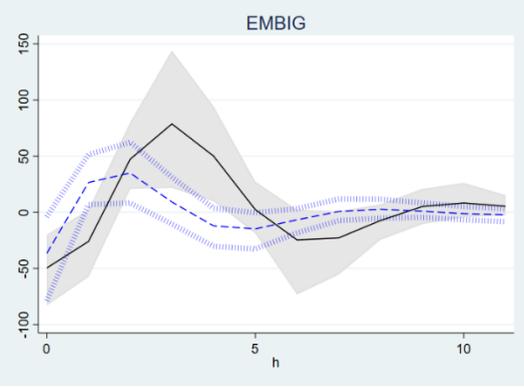

Notes: See notes to Figure 11.

# Figure 14: r: Nominal Exchange Rate for US

A: $r \to p$          B: $r \to y$          C: $r \to p \to y$

## CTOT1

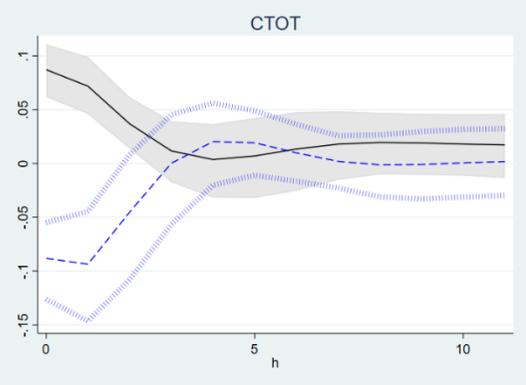 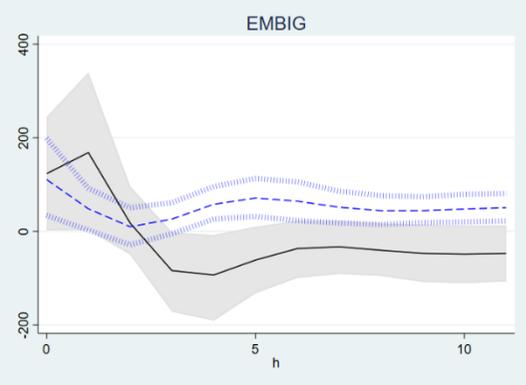 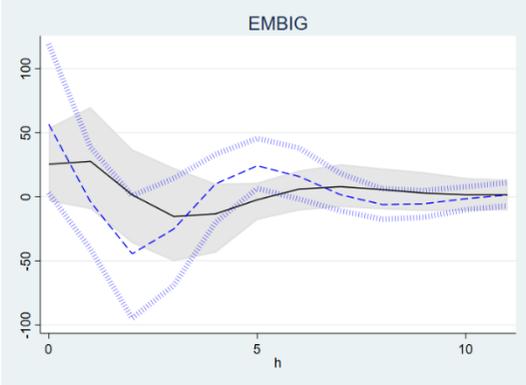

## CTOT2

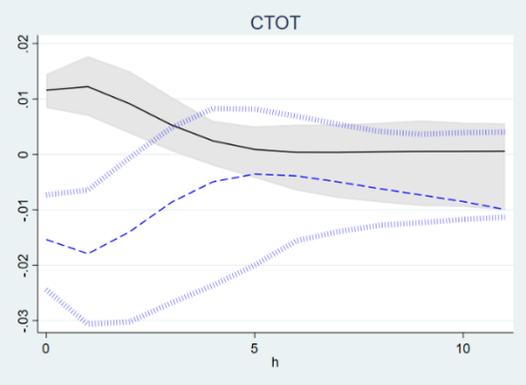 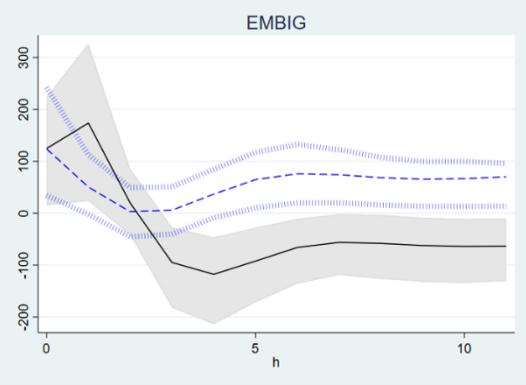 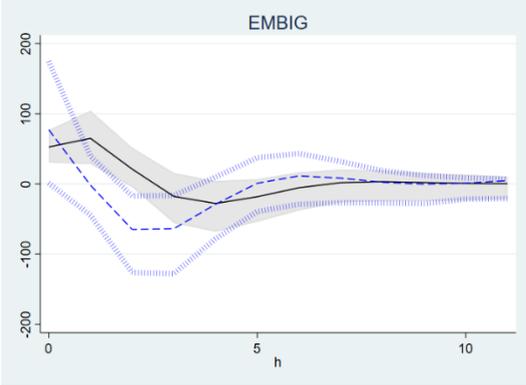

Notes: See notes to Figure 11.